\documentclass[preprint,groupedaddress, showpacs, floatfix]{revtex4-1}
\usepackage{graphicx}

\begin{document}

\title{Boron and nitrogen impurities in SiC nanoribbons: an {\it ab initio} investigation}
\author{C. D. Costa}
\author{J. M. Morbec}
\email{jmmorbec@gmail.com}
\affiliation{Instituto de Ci\^encias Exatas, Universidade Federal de Alfenas, CEP 37130-000,
Alfenas, MG, Brazil}

\date{\today}

\begin{abstract}
Using {\it ab initio} calculations based on density-functional theory we have performed a theoretical 
investigation of substitutional boron and nitrogen impurities in silicon carbide (SiC) nanoribbons. We have considered  
hydrogen terminated SiC ribbons with zigzag and armchair edges. 
In both systems we verify that the boron and nitrogen atoms energetically prefer to be localized at the edges of the nanoribbons. 
However, while boron preferentially substitutes a silicon atom, nitrogen prefers to occupy a carbon site. 
In addition, our electronic-structure calculations 
indicate that (i) substitutional boron and nitrogen impurities do not affect the semiconducting character of the armchair SiC nanoribbons, 
and (ii) the half-metallic behavior of the zigzag nanoribbons is maintained in the presence of substitutional boron impurities.  
In contrast, nitrogen atoms occupying edge carbon sites transform half-metallic zigzag nanoribbons into metallic systems.
\end{abstract}



\maketitle

\section{Introduction} \label{intro}

Silicon carbide (SiC) nanostructures have attracted much interest in the last years because 
they combine the excellent properties of SiC bulk\cite{choyke, madar, masri} (wide band gap, high thermal conductivity, 
high breakdown electric field, thermal stability, high electronic mobility, and others) with quantum-size effects. This combination 
makes SiC nanostructures interesting materials 
for nanotechnology applications and nanoscale engineering.\cite{melinon} 

Recently, increasing attention has been paid to the planar structures of SiC. 
Although silicon prefers $sp^3$ instead of $sp^2$ hybridization, 
SiC nanoribons (SiCNRs) have been successfully synthesized\cite{h-zhang} and 
first-principles calculations have shown that SiC has a stable two-dimensional honeycomb structure, similar to graphene.\cite{h-sahin, yu, beka} 
In addition, recent {\it ab initio} studies revealed that the monolayer SiC sheet is semiconducting with a band gap of about 2.55~eV,\cite{h-sahin, yu, beka} 
while armchair SiCNRs are nonmagnetic semiconductors 
(with a wide band gap for all widths),\cite{lian-sun, j-zhang} and zigzag SiCNRs narrower 4~nm present half-metallic behavior 
without applied external electric field or chemical modifications.\cite{lian-sun, j-zhang, lou} This feature makes 
narrow zigzag SiCNRs promising candidates for spintronic applications. 

The great potential of the SiC nanoribbons for future applications in spintronics stimulates detailed studies of the  
effects of impurities and defects in their electronic properties.  
In particular, SiC nanoribbons doped with boron and nitrogen atoms is an important issue to be investigated, 
since B and N are common contaminants in SiC bulk, acting as p- and n-type dopants, respectively. 
Moreover, it is well known that substitutional 
boron and nitrogen impurities in zigzag graphene nanoribbons break the spin degeneracy  
of the transmittance channels, and transform metallic nanoribbons into semiconducting systems.\cite{martins, martins1, rocha, sodi}

In this work we have performed a theoretical investigation, using {\it ab initio} calculations based on the density-functional theory, 
of substitutional boron and nitrogen impurities in armchair and zigzag SiC nanoribbons. 
Our results indicate that the boron and nitrogen atoms energetically prefer to be localized at the edges of both nanoribbons. 
However, while boron preferentially substitutes 
a silicon atom, nitrogen prefers to occupy a carbon site. In addition, we verify that 
substitutional B and N impurities do not affect the semiconducting character of the 
armchair SiCNRs, although the spin degeneracy of the band structure is broken. 
On the other hand, we observe different effects of these impurities on the half-metallic behavior of the zigzag SiCNRs: 
the half-metallicity of the nanoribbons is maintained in the presence of substitutional B impurities, 
whereas zigzag SiCNRs become metallic when doped with N atoms.

\section{Method of calculation} \label{method}

The {\it ab initio} calculations presented in this work were carried out within the framework of the 
density functional theory (DFT),\cite{dft} using the SIESTA code.\cite{siesta}
We employed the generalized gradient approximation as implemented by Perdew, Burke and Ernzerhof (GGA-PBE)\cite{pbe} 
for the exchange-correlation functional, and 
norm-conserving fully-separable pseudopotentials\cite{pseudo} 
to treat the electron-ion interactions. The Kohn-Sham orbitals were expanded in a linear combination of 
numerical pseudoatomic orbitals\cite{lcao} using a split-valence double-zeta basis set with  polarization functions (DZP).\cite{dzp}

\begin{figure}
\includegraphics[scale=0.55]{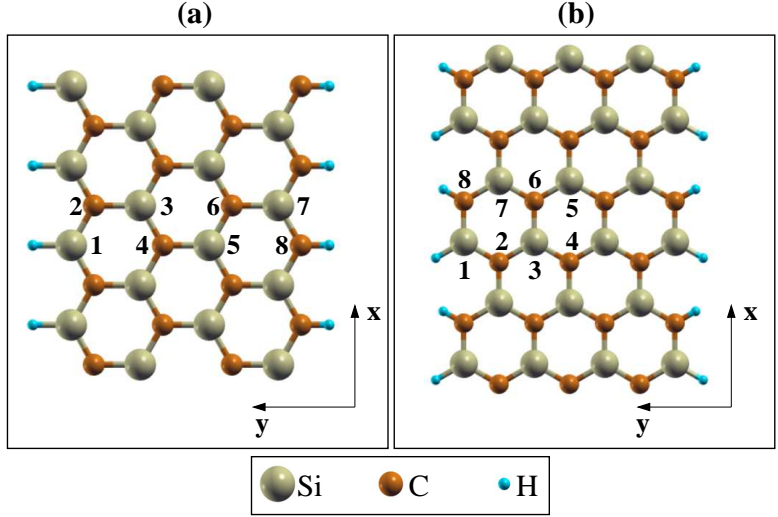}
\caption{\label{fig1}
(Color online) Structural models of (a) 4-ZSiCNR and (b) 7-ASiCNR indicating the 
substitutional sites investigated. Odd (even) numbers correspond to Si (C) sites.}
\end{figure}

We have considered hydrogen-passivated zigzag and armchair SiCNRs with widths $W=4$ (4-ZSiCNR) and $W=7$ (7-ASiCNR),\cite{footnote1} respectively.  
The 4-ZSiCNR and 7-ASiCNR structures were modeled within the supercell approach with 40 and 54 atoms (including hydrogen atoms) in the unit cell, respectively.  
The structural models of 4-ZSiCNR and 7-ASiCNR, indicating the substitutional sites investigated, are shown in  Fig.~\ref{fig1}. 
A vacuum region of about 20 \AA \ along the nonperiodic directions was employed to avoid interactions between two neighboring ribbons. 
All the atomic positions were relaxed during the geometry optimization until the Hellman-Feynman forces were below 0.01~eV/\AA. 
The Brillouin zone was sampled using the Monkhorst-Pack scheme\cite{mp} with a $(11 \times 1 \times 1)$ grid for the total-energy calculations 
and a $(120 \times 1 \times 1)$ mesh for the electronic-structure calculations. The convergence of our total-energy results with respect to the number of 
{\bf k}-points was verified.

In each system the energetic stability of the impurities was determined by comparing their formation energies\cite{sabi, northup}
\begin{equation}
\Omega=E_{T}-\sum_{i}{N_i\mu_i},
\label{eq-form-ener}
\end{equation}
where $E_{T}$ is the total energy of the structure, $\mu_i$ is the chemical potential of the atomic specie $i$, and 
$N_i$ is the number of atoms $i$ in the structure. 
The chemical potentials of Si and C are restricted to the ranges  
\begin{equation}
-\Delta H({\rm SiC})+\mu_{\rm Si}^{\rm bulk} \leq \mu_{\rm Si} \leq \mu_{\rm Si}^{\rm bulk}
\label{limit1}
\end{equation}
and
\begin{equation}
-\Delta H({\rm SiC})+\mu_{\rm C}^{\rm bulk} \leq \mu_{\rm C} \leq \mu_{\rm C}^{\rm bulk}, 
\label{limit2}
\end{equation}
where $\Delta H({\rm SiC})$ is the SiC formation heat.

The (relative) formation energies presented in this work were calculated assuming Si/C stoichiometric condition 
\begin{equation}
\mu_{\rm C, Si}=\mu_{\rm C, Si}^{\rm bulk}-\frac{1}{2}\Delta H({\rm SiC}). 
\label{eq-stoi}
\end{equation}

\section{Results and discussions} \label{results}

\begin{figure}
\includegraphics[scale=0.35]{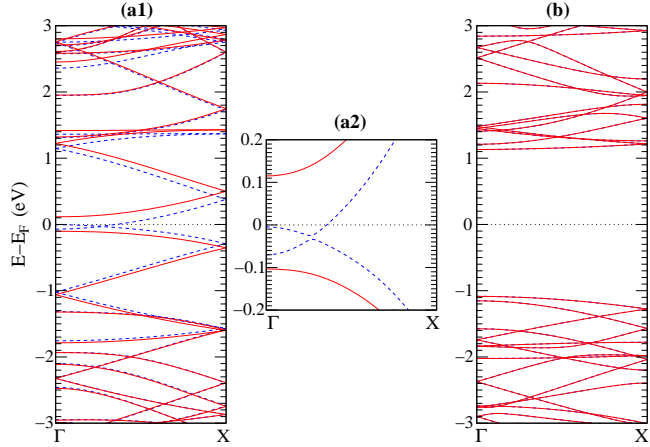}
\caption{\label{fig2}
(Color online) Electronic band structures of pristine (a) 4-ZSiCNR and (b) 7-ASiCNR. 
The band structure of pristine 4-ZSiCNR is depicted in two different ranges: (a1)~$|E-E_F| \leq 3.0$~eV and 
(a2) $|E-E_F| \leq 0.2$~eV. 
Spin-up and spin-down bands are indicated by solid red and 
dashed blue lines, respectively. 
Dotted line indicates the position of the Fermi level.}
\end{figure}

Initially we examined the equilibrium geometries and the electronic structures  
of pristine 4-ZSiCNR and 7-ASiCNR. 
We found optimized Si-C bond lengths from 1.77 to 1.84 \AA \ in 4-ZSiCNR, and from 1.75 to 1.81~\AA \ in 7-ASiCNR. 
In both systems, the Si-H and C-H bond lengths were about 1.52 and 1.11~\AA, 
respectively. These values are 
in good agreement with previous calculations for SiC nanoribbons,\cite{j-zhang} sheets\cite{h-sahin} and nanotubes.\cite{menon, xia}
By analysing the electronic structure in Fig.~\ref{fig2} we observed that pristine 4-ZSiCNR [Fig.~\ref{fig2}~(a)] 
presents a half-metallic behavior, with a semiconducting spin-up channel (band gap of 0.22~eV at the $\Gamma$ point) and a metallic spin-down channel. 
On the other hand, pristine 7-ASiCNR [Fig.~\ref{fig2}~(b)] is a semiconductor with a direct band gap of 2.22~eV (at the $\Gamma$ point)  
for both spin-up and spin-down channels. 

Considering one B or N atom per supercell we investigated 8 substitutional sites 
(sites labeled from 1 to 8 in Fig.~\ref{fig1}) in both 4-ZSiCNR and 7-ASiCNR. 
In B-doped nanoribbons (B/4-ZSiCNR and B/7-ASiCNR) the formation energy of each B$_{\rm A}^i$ configuration was compared with that of 
B$_{\rm Si}^1$ configuration 
\begin{equation}
\Delta \Omega_{\rm B} = \Omega({\rm B}_{\rm A}^{i})-\Omega({\rm B}_{\rm Si}^{1}).
\label{b-energy}
\end{equation} 
Here $i\in \{1, ..., 8\}$, ${\rm A} \in \{{\rm Si}, {\rm C}\}$ and B$_{\rm Si}^1$ denotes the configuration in which B substitutes Si 
at the site 1. 
In the N/4-ZSiCNR and N/7-ASiCNR systems, we compared each 
N$_{\rm A}^i$ configuration with N$_{\rm C}^8$ configuration using 
\begin{equation}
\Delta \Omega_{\rm N} = \Omega({\rm N}_{\rm A}^{i})-\Omega({\rm N}_{\rm C}^{8}).
\label{n-energy}
\end{equation} 
Our results for the relative formation energies  
$\Delta \Omega_{\rm B}$ 
and $\Delta \Omega_{\rm N}$ 
are summarized in Table~\ref{tab1}. 
In this case, positive values of $\Delta \Omega_{\rm B}$ ($\Delta \Omega_{\rm N}$) indicate that B$_{\rm Si}^1$ (N$_{\rm C}^8$) 
is energetically more favorable than  B$_{\rm A}^i$ (N$_{\rm A}^i$). 

\begin{table}
\caption{Relative formation energies (in eV) 
of substitutional B ($\Delta \Omega_{\rm B}$) and N ($\Delta \Omega_{\rm N}$) impurities in 4-ZSiCNR and 7-ASiCNR. 
$\Delta \Omega_{\rm B}$  
and $\Delta \Omega_{\rm N}$ 
are given by Eqs.~(\ref{b-energy}) and (\ref{n-energy}). 
The occupation sites are indicated in Fig.~\ref{fig1}. }
\begin{ruledtabular}
\begin{tabular}{lcccccc}
& & \multicolumn{2}{c}{4-ZSiCNR} & & \multicolumn{2}{c}{7-ASiCNR} \\
\cline{3-4} \cline{6-7}
Sites  & & B & N & & B  & N \\
\hline
Si$_1$ & & 0.00  & 1.90 & & 0.00  & 0.85\\ 
C$_2$ & &  1.16  & 1.35 & & 1.07  & 0.39\\
Si$_3$ & & 0.61  & 3.88 & & 0.48  & 3.19\\
C$_4$ & & 2.01   & 1.37 & & 1.29  & 1.11\\ 
Si$_5$ & & 1.24  & 4.30 & & 0.59  & 3.42\\
C$_6$ & &  2.30  & 0.74 & & 1.31  & 1.08\\
Si$_7$ & & 1.22  & 2.61 & & 0.29  & 2.34\\
C$_8$ & & 1.95   & 0.00 & & 0.89  & 0.00\\
\end{tabular}
\end{ruledtabular}
\label{tab1}
\end{table}

As can be seen in Table I, B$_{\rm Si}^1$ and N$_{\rm C}^8$ are the 
most favorable configurations in both 4-ZSiCNR and 7-ASiCNR. This indicates that the B and N atoms energetically prefer to be 
localized at the egde of the nanoribbons. However, while B preferentially occupies a Si site, N prefers to substitute a C atom. 
Among the B$_{\rm C}$ configurations, B$_{\rm C}^2$ and B$_{\rm C}^8$ are the most stable in 4-ZSiCNR and 7-ASiCNR, respectively. 
However, at the Si/C stoichiometric condition, B$_{\rm C}^2$ (B$_{\rm C}^8$) is energetically less favorable than B$_{\rm Si}^1$ by 1.16 (0.89)~eV   
in 4-ZSiCNR (7-ASiCNR). Extending the calculations of the formation energies $\Omega({\rm B}_{\rm C}^{2})$, $\Omega({\rm B}_{\rm C}^{8})$ and  
$\Omega({\rm B}_{\rm Si}^{1})$ for different values of $\mu_{\rm C}$ and $\mu_{\rm Si}$, we verify that   
$\Omega({\rm B}_{\rm C}^{2}) > \Omega({\rm B}_{\rm Si}^{1})$ and 
$\Omega({\rm B}_{\rm C}^{8}) > \Omega({\rm B}_{\rm Si}^{1})$ 
within the limits given by the expressions (\ref{limit1}) and (\ref{limit2}) with $\Delta H({\rm SiC})=0.72$~eV.\cite{footnote2} 
These results indicate that 
B$_{\rm C}$ is not expected to occur in both armchair and zigzag SiCNRs. 
Performing the same analysis for N impurities, we observe that 
N$_{\rm Si}^1$ is the most likely configuration among the N$_{\rm Si}$ configurations in both 4-ZSiCNR and 7-ASiCNR, but  
N$_{\rm Si}^1$ is less favorable than N$_{\rm C}^8$ within the allowed ranges for $\mu_{\rm C}$ and $\mu_{\rm Si}$ 
(expressions (\ref{limit1}) and (\ref{limit2})). Thus, it is not expected that N atoms occupy Si sites in SiC nanoribbons. 

The energetic preference of the B (N) atoms for Si (C) sites at the stoichiometric condition  
has been also observed in SiC nanotubes\cite{gali} and nanowires.\cite{igor} 
In these nanostructures, like in SiC nanoribbons, the formation of N$_{\rm Si}$ is not expected to occur.
Nevertheless, in SiC nanotubes and nanowires B$_{\rm C}$ becomes 
more likely than B$_{\rm Si}$ at the limit $\mu_{\rm Si}= \mu_{\rm Si}^{\rm bulk}$ 
and $\mu_{\rm C}=\mu_{\rm C}^{\rm bulk}-\Delta H({\rm SiC})$.\cite{gali, igor}

\begin{figure}
\includegraphics[scale=0.50]{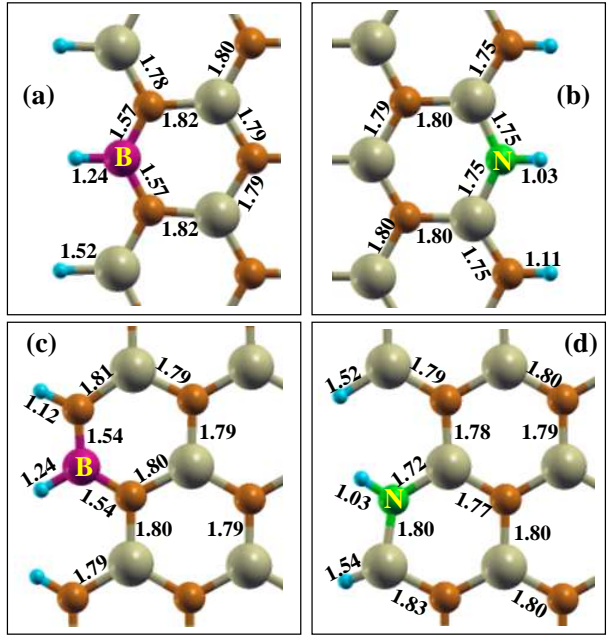}
\caption{\label{fig3}
(Color online) Equilibrium geometries of (a) B$_{\rm Si}^1$/4-ZSiCNR, (b) N$_{\rm C}^8$/4-ZSiCNR, 
(c) B$_{\rm Si}^1$/7-ASiCNR, and (d) N$_{\rm C}^8$/7-ASiCNR. 
The atomic distances are in \AA.}
\end{figure}

We next examined the equilibrium geometries of the most likely configurations (B$_{\rm Si}^1$ and N$_{\rm C}^8$) in 
4-ZSiCNR and 7-ZSiCNR. The variations in the Si-C bond lengths 
with respect to the pristine structures were less than 0.03~\AA \ 
in the B$_{\rm Si}^1$/4-ZSiCNR,  N$_{\rm C}^8$/4-ZSiCNR and B$_{\rm Si}^1$/7-ASiCNR systems, and less than 0.05~\AA \ 
in  N$_{\rm C}^8$/7-ASiCNR. As depicted in Fig.~\ref{fig3}, we found 
(i) B-C and B-H bond lengths of 1.57 and 1.24~\AA \ (1.54 and 1.24 \AA), respectively, 
in the  B$_{\rm Si}^1$/4-ZSiCNR (B$_{\rm Si}^1$/7-ASiCNR) system, and (ii) N-Si and N-H bond lengths of 
1.75 and 1.03~\AA, respectively, in  N$_{\rm C}^8$/4-ZSiCNR. In the N$_{\rm C}^8$/7-ASiCNR structure, besides the N-Si bond lengths of 1.72 and 
1.80~\AA, we also observed that the Si$_1$ atom [see Fig.~\ref{fig1}(b)] 
moves upward (along the $z$ direction) by 0.28~\AA \ with respect to the pristine 7-ASiCNR, 
while the hydrogen atom bonded to Si$_1$ is displaced by 0.77~\AA \ in the opposite direction. 
A similar behavior is found in N$_{\rm C}^2$/7-ASiCNR (the second more stable configuration for N/7-ASiCNR), 
where the Si$_1$ and H atoms are displaced by 0.25 and 0.87~\AA, respectively. 
By constraining the relaxation of the N$_{\rm C}^8$/7-ASiCNR and N$_{\rm C}^2$/7-ASiCNR systems 
along the $z$ direction (the atoms are free to relax 
in the $x$ and $y$ directions), we verify that their total energies are increased by 0.60 and 0.62~eV, respectively, 
in comparison with the fully relaxed structures. 
Even so, N$_{\rm C}^8$ remains the most favorable configuration for N/7-ASiCNR.

\begin{figure}
\includegraphics[scale=0.40]{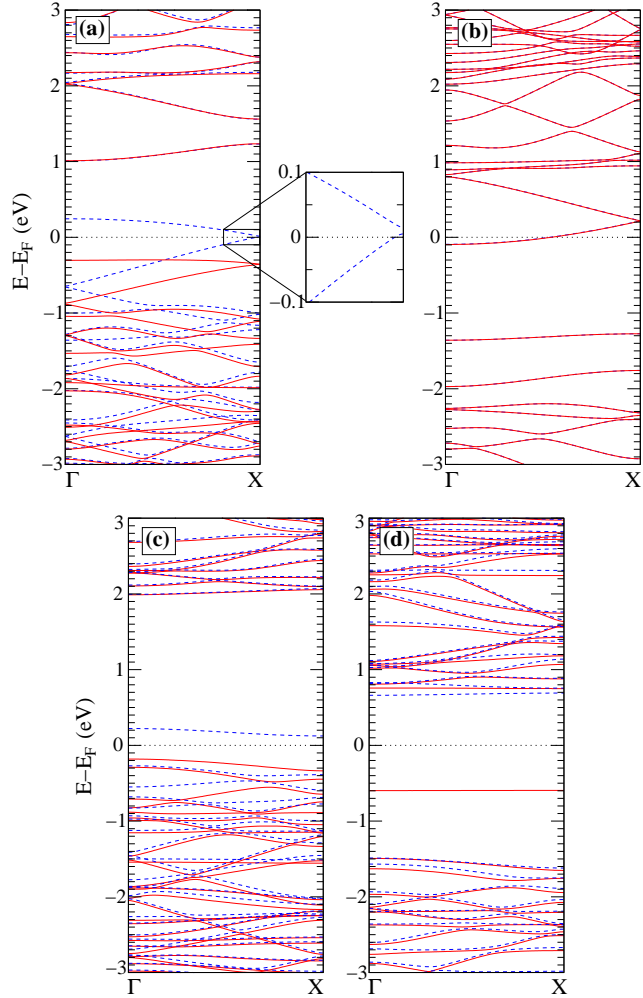}
\caption{\label{fig4}
(Color online) Electronic band structures of  (a) B$_{\rm Si}^1$/4-ZSiCNR, (b) N$_{\rm C}^8$/4-ZSiCNR, 
(c) B$_{\rm Si}^1$/7-ASiCNR, and (d) N$_{\rm C}^8$/7-ASiCNR. Spin-up and spin-down bands are indicated by solid red and 
dashed blue lines, respectively. 
Dotted line indicates the position of the Fermi level.}
\end{figure}

\begin{figure}
\includegraphics[scale=0.40]{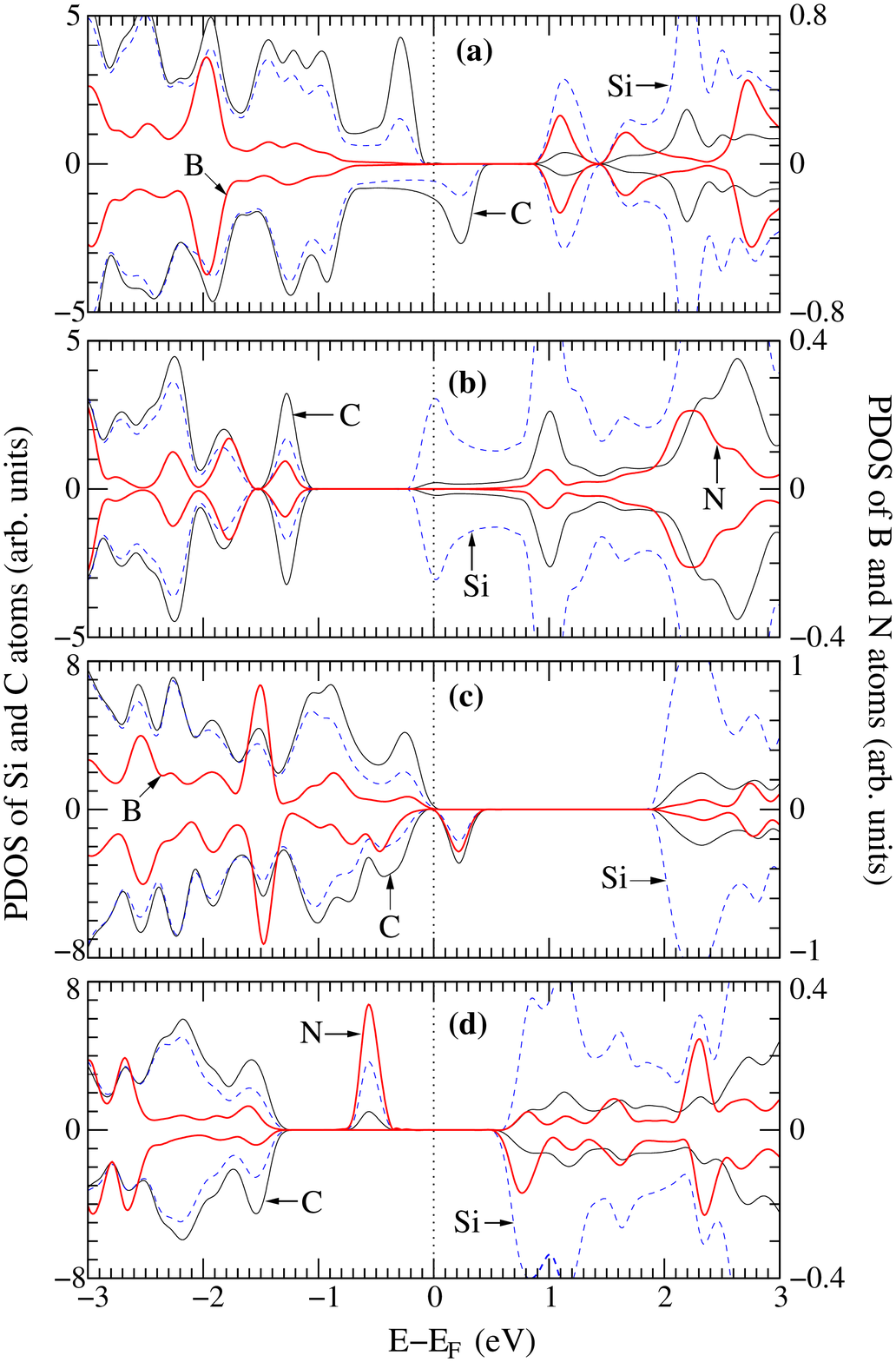}
\caption{\label{fig5}
(Color online) Projected density of states (PDOS) of (a) B$_{\rm Si}^1$/4-ZSiCNR, (b) N$_{\rm C}^8$/4-ZSiCNR, 
(c) B$_{\rm Si}^1$/7-ASiCNR, and (d) N$_{\rm C}^8$/7-ASiCNR. Solid thick red lines indicate the PDOS of the B or N atoms.  
Dashed blue and solid (thin) black lines represent the PDOS of the Si and C atoms, respectively. 
The position of the Fermi level is indicated by dotted lines. 
Positive and negative values of PDOS correspond to spin-up and spin-down contributions, respectively. Left vertical scale: PDOS of Si and C atoms. 
Right vertical scale: PDOS of B or N atoms.}
\end{figure}

Figures \ref{fig4} and  \ref{fig5} display the electronic band structures and the projected density of states (PDOS) 
of B$_{\rm Si}^1$/4-ZSiCNR, N$_{\rm C}^8$/4-ZSiCNR, 
B$_{\rm Si}^1$/7-ASiCNR, and N$_{\rm C}^8$/7-ASiCNR. In Fig.~\ref{fig4}(a) we observe that 
the B$_{\rm Si}^1$/4-ZSiCNR system presents half-metallic characteristics: the spin-up channel is semiconducting, 
with a direct band gap of 1.30~eV at the $\Gamma$ point, and the spin-down channel is  metallic, with an energy band crossing the 
Fermi level ($E_{F}$) near to the $X$ point. 
This result indicates that one substitutional B impurity per supercell 
does not affect the half-metallic behavior of 4-ZSiCNR. 
The PDOS of the B$_{\rm Si}^1$/4-ZSiCNR system [Fig.~\ref{fig5}(a)] reveals that 
the partially occupied spin-down electronic state, as well as the highest-occupied spin-up state, 
is mostly composed by C~$2p$ and Si~$3d$ orbitals, without contributions from B$_{\rm Si}^1$. 
The B~$2p$ orbitals are localized below $(E_{F}-0.5)$~eV and above $(E_{F}+0.9)$~eV. 

As depicted in Figs.~\ref{fig4}(b) and \ref{fig5}(b), the N$_{\rm C}^8$/4-ZSiCNR system is metallic, with degenerate 
spin-up and spin-down channels. The partially occupied state comes mainly from the Si~$3p$ orbitals 
localized along the zigzag edge opposite to N$_{\rm C}^8$. 
The occupied state situated between 1.2 and 1.4~eV below the Fermi level is mostly composed by Si~$3p$, Si~$3d$, C~$2p$ and N~$2p$ orbitals.

Figures \ref{fig4}(c) and \ref{fig4}(d) show that the semiconducting character of pristine 7-ASiCNR is maintained in the presence of substitutional 
B and N impurities, but the spin degeneracy of the band structure is broken. The B$_{\rm Si}^1$/7-ASiCNR system [Fig.~\ref{fig4}(c)] 
presents a direct band gap of 2.17~eV (at $\Gamma$ point) for the spin-up channel and an indirect band gap of 0.39~eV 
for the spin-down channel. We found an unoccupied spin-down state lying between 0.1 and 0.3~eV above the Fermi level. This state consists 
mainly of  B~$2p$, C~$2p$, Si~$3p$ and Si~$3d$ orbitals [Fig.~\ref{fig5}(c)]. In N$_{\rm C}^8$/7-ASiCNR [Fig.~\ref{fig4}(d)], 
both spin-up and spin-down channels exhibit semiconducting characteristics, with direct band gaps of 1.34~eV (at the X point) and 
2.15~eV (at the $\Gamma$ point), respectively. The highest-occupied spin-up state is very localized and lies at around 0.6~eV below the Fermi level. 
The PDOS of N$_{\rm C}^8$/7-ASiCNR [Fig.~\ref{fig5}(d)] indicates that this state is mainly composed by Si $3p$, Si $3d$ and N $2p$ orbitals, 
and by the $2p$ orbitals of the inner C atoms. We verified that the C~$2p$ orbitals localized along the edges do not contribute to this electronic 
state. 

\section{Conclusions} \label{conclusion}

In summary, we performed a theoretical investigation, using {\it ab initio} calculations, 
of substitutional B and N impurities in SiC nanoribbons.  
In both ASiCNR and ZSiCNR we found an energetic preference for B and N atoms occupying edge sites. 
However, we verified that B preferentially substitutes a Si atom, whereas N prefers to occupy a C site. 
We also observed that the formation of B$_{\rm C}$ and N$_{\rm Si}$ is not expected to occur in SiCNRs. 
In addition, our electronic-structure calculations revealed that B and N impurities have different effects on the 
electronic character of SiCNRs. We observed that (i) substitutional B and N impurities do not affect the 
semiconducting behavior of ASiCNRs, and (ii) the half-metallicity of ZSiCNRs is maintained in the presence of substitutional B atoms. 
In contrast, we verified that half-metallic ZSiCNRs become metallic when doped with N atoms. 
These results suggest that the electronic properties of SiCNRs can be controlled by B- and N-doping processes.

\begin{acknowledgments}

The authors acknowledge the financial support from the Brazilian agency FAPEMIG and the computational facilities of CENAPAD/SP. 
J.M.M. would like to thank Gul Rahman and Roberto Hiroki Miwa for fruitful discussions.

\end{acknowledgments}


\begin{thebibliography}{99}

\bibitem{choyke} W.~J.~Choyke, H.~Matsunami, and G.~Pensl (eds), {\it Silicon Carbide: Recent Major
Advances} (Springer, Berlin, 2004).

\bibitem{madar} R.~Madar, Nature {\bf 430}, 974 (2004).

\bibitem{masri} P.~Masri, Surf. Sci. Rep. {\bf 48}, 1 (2002).

\bibitem{melinon} P.~M\'elinon, B.~Masenelli, F.~Tournus, and A.~Perez, Nature Mater. {\bf 6}, 479 (2007).

\bibitem{h-zhang} H.~Zhang, W.~Ding, K.~He, and M.~Li, Nanoscale Res. Lett. {\bf 5}, 1264 (2010).

\bibitem{h-sahin} H.~Sahin, S.~Cahangirov, M.~Topsakal, E.~Bekaroglu, E.~Akturk, R.~T.~Senger, and S.~Ciraci, 
Phys. Rev. B {\bf 80}, 155453 (2009).

\bibitem{yu} M.~Yu, C.~S.~Jayanthi, and S.~Y.~Wu, Phys. Rev. B {\bf 82}, 075407 (2010).

\bibitem{beka} E.~Bekaroglu, M.~Topsakal, S.~Cahangirov, and S.~Ciraci, Phys. Rev. B {\bf 81}, 075433 (2010).

\bibitem{lian-sun} L.~Sun, Y.~Li, Z.~Li, Q.~Li, Z.~Zhou, Z.~Chen, J.~Yang, and J.~G.~Hou, J. Chem. Phys. {\bf 129}, 174114 (2008).

\bibitem{j-zhang} J.-M.~Zhang, F.-L.~Zheng, Y.~Zhang, and V.~Ji, J. Mater. Sci. {\bf 45}, 3259 (2010).

\bibitem{lou} P.~Lou and J.~Y.~Lee, J. Phys. Chem. C {\bf 113}, 12637 (2009).

\bibitem{martins} T.~B.~Martins, R.~H.~Miwa, A.~J.~R.~Silva, and A.~Fazzio, Phys. Rev. Lett. {\bf 98}, 196803 (2007).

\bibitem{martins1} T.~B.~Martins, A.~J.~R.~Silva, R.~H.~Miwa, and A.~Fazzio, Nano Lett. {\bf 8}, 2293 (2008). 

\bibitem{rocha} A.~R.~Rocha, T.~B.~Martins, A.~Fazzio, and A.~J.~R.~Silva, Nanotechnology {\bf 21}, 345202 (2010).

\bibitem{sodi} F.~C.-Sodi, G.~Cs\'anyi, S.~Piscanec, and A.~C.~Ferrari, Phys. Rev. B {\bf 77}, 165427 (2008).

\bibitem{dft} P.~Hohenberg and W.~Kohn, Phys. Rev. {\bf 136}, B864 (1964). 

\bibitem{siesta} J.~M.~Soler, E.~Artacho, J.~D.~Gale, A.~Garc\'{i}a, J.~Junquera, P.~Ordej\'on, and D.~S\'anchez-Portal, 
J. Phys.: Condens. Matter {\bf 14}, 2745 (2002). 

\bibitem{pbe} J.~P.~Perdew, K.~Burke, and M.~Ernzerhof, Phys. Rev. Lett. {\bf 77}, 3865 (1996). 

\bibitem{pseudo} N.~Troullier and J.~L.~Martins, Phys. Rev. B {\bf 43}, 1993 (1991).

\bibitem{lcao} O.~F.~Sankey and D.~J.~Niklewski, Phys. Rev. B {\bf 40}, 3979 (1989).

\bibitem{dzp} E.~Artacho, D.~S\'anchez-Portal, P.~Ord\'ejon, A.~Garcia, and J.~M.~Soler, Phys. Status Solidi B {\bf 215}, 809 (1999).

\bibitem{footnote1} According to Refs.~\onlinecite{lian-sun, j-zhang} the width $W$ of a SiCNR is defined as the 
number of zigzag chains across the ribbon width, for ZSiCNR, or the number of dimer lines for ASiCNRs.

\bibitem{mp} H.~J.~Monkhorst and J.~D.~Pack, Phys. Rev. B {\bf 13}, 5188 (1976).

\bibitem{sabi} M.~Sabisch, P.~Kr\"uger, and J.~Pollmann, Phys. Rev. B {\bf 55}, 10561 (1997).

\bibitem{northup} J.~E.~Northrup and S.~B.~Zhang, Phys. Rev. B {\bf 47}, 6791 (1993).

\bibitem{menon} M.~Menon, E.~Richter, A.~Mavrandonakis, G.~Froudakis, and A.~N.~Andriotis, Phys. Rev. B {\bf 69}, 115322 (2004).

\bibitem{xia} M.~W.~Zhao, Y.~Xia, R.~Q.~Zhang, and S.-T.~Lee, J. Chem. Phys. {\bf 122}, 214707 (2005).

\bibitem{footnote2} Note that the calculations of $\Delta \Omega_{\rm B}$ and $\Delta \Omega_{\rm N}$ at the Si/C stoichiometric condition 
do not depend on the value of $\Delta H({\rm SiC})$. For example: using Eqs.~(\ref{eq-form-ener}) and (\ref{eq-stoi}) 
we obtain $\Delta \Omega_B = \Omega({\rm B}_{\rm C}^2)-\Omega({\rm B}_{\rm Si}^1)
=E_T({\rm B}_{\rm C}^2)-E_T({\rm B}_{\rm Si}^1)+\mu_{\rm C}^{\rm bulk}-\mu_{\rm Si}^{\rm bulk}$ at the stoichiometric 
condition.

\bibitem{gali} A.~Gali, Phys. Rev. B {\bf 73}, 245415 (2006).

\bibitem{igor} I.~S.~S.~Oliveira and R.~H.~Miwa, Phys. Rev. B {\bf 79}, 085427 (2009). 

\end{thebibliography}
\end{document}